\documentclass{kapproc} 
\usepackage{procps} 
\usepackage[dvips]{graphicx}
\upperandlowercase
\kluwerbib 

\begin{document}

\articletitle[Local Starbursts: Perspectives from the Optical]{LOCAL STARBURSTS: \\ PERSPECTIVES FROM THE OPTICAL}

\author{Daniela Calzetti}

 \affil{Space Telescope Science Institute\\
3700 San Martin Drive, Baltimore, MD 21218, U.S.A.}
 \email{calzetti@stsci.edu}

\begin{abstract}
The optical regime is historically the best-studied wavelength range.
Gas ionized by massive stars produces optical emission lines that have
been used to derive indicators of star--formation rate, metallicity,
dust reddening, and the ionization conditions of the interstellar
medium. Absorptions lines have been used to measure velocity
dispersions, and the 4000~\AA\ break has been shown to be a useful
indicator of the mean age of stellar populations. I briefly summarize
some recent work done on, specifically, star formation rate
indicators, in view of their importance for understanding
star--forming galaxies at high redshift.
\end{abstract}


\section{Introduction}

In recent years, the advent of improved infrared instrumentation on large,
8--10m class, telescopes has opened a window on rest-frame optical
observations of high redshift galaxies, and revived interest for this
historically well--studied wavelength regime.  

The easy access from the ground to the optical emission from local
celestial bodies (stars, HII regions, galaxies, etc.) has led to the
development and definition of a series of tracers of the physical and
chemical conditions of these objects. Of immediate relevance for
complex systems like galaxies are the nebular emission lines, stellar
and interstellar absorption lines, and broad--band features like the
4000~\AA\ break.

The 4000~\AA\ break, for instance, is an estimator of a stellar
population's mean age (e.g., Kauffmann et al. 2003), while absorption
lines have been widely employed to measure velocity dispersions within
the stellar systems. Among all optical tracers of physical and
chemical conditions, however, the lion's share goes to the nebular
emission lines. The gas ionized by young, massive stars produces
optical emission lines from a number of chemical elements, and with a
fairly large range of intensities; these have been `calibrated' to
`measure':
\begin{itemize}
\item star formation rates (SFR; from [OII], H$\alpha$, ...);
\item gas chemical abundances (O, N, S, ...);  
\item dust reddening (e.g., from the Balmer series);
\item diagnostics of star--formation feedback, and, in general,
ionization conditions ([OI], [NII], [SII], ...).
\end{itemize}

The applicability of any such indicator for investigations at
cosmological distances depends on the redshift range under
consideration, and the number of lines that can be accessed. The
[OII]($\lambda$3727~\AA) emission line is sufficiently blue to be
observable in the optical window up to redshift less or about 1.5;
however, this line alone (with no other information at shorter or
longer wavelengths) can only provide a highly uncertain estimate of the
distant galaxies' SFRs (see next section, and, e.g., Hammer et~al.\ 
1997, Hogg et~al.\ 1998, Rosa--Gonzalez et~al.\ 2002, Hippelein et~al.\ 
2003, Teplitz et~al.\ 2003, Kewley et~al.\ 2004). New infrared
instruments are providing more leverage, by allowing investigators to
access a larger suite of restframe optical emission lines, from
[OII]($\lambda$3727~\AA) to [NII]($\lambda$6584~\AA), up to, for some
lines, redshift z$\sim$3. Multiple emission lines from the same
cosmological object afford better estimates of dust reddening,
gas chemical abundances, etc.\ (e.g., Teplitz et~al.\ 2000, Pettini et~al.\ 
2001, Lemoine--Busserolle et~al.\ 2003). Last, but not least, once
the James Webb Space Telescope is on orbit, from its vantage point
above the atmosphere it will provide an unobstructed view of the
earliest galaxies, detecting H$\alpha$ up to z$\approx$6.5, and,
potentially, as high as z$\approx$40 (depending on sensitivity and
whether ionizing objects exist at such high redshifts). Redshift z=6.5
corresponds to an epoch when the Universe was 6\% of its current age
(for a $\Lambda$CDM cosmology).

Given that the high--redshift `frontier' employs tools derived from
the more accessible low--redshift Universe to understand the evolution
of galaxies, it is worth revisiting the strengths and limitations of
some of these tools, and whether more investigation is needed in some
areas. For instance, recent studies have re-iterated the limitations
of the well--known and well--established `strong line method' for
chemical abundance measurements in metal--rich environments (Garnett
et~al.\ 2004).

In this talk I concentrate on the SFR indicators accessible in the
optical regime, highlighting recent progress in the area; I connect
these optical indicators to those at other wavelengths, suggesting
where additional investigation may be needed.

\section{Star Formation Rate Indicators in the Optical}

The basic questions that come to mind when using SFR indicators at
optical or other wavelengths are: are they consistent with one
another? What level of `uncertainty' each of them carries, and what
factors produce such uncertainty?

At optical wavelengths, the two most widely employed SFR tracers are:
the Balmer lines emission (H$\alpha$, H$\beta$, ...) and the
[OII]($\lambda$3727~\AA) doublet line emission. Both are measures of
`instantaneous' star formation, as the gas is excited by the ionizing
photons of the short--lived O and early--B stars. The intensity of
both Balmer and [OII] lines is affected by dust extinction -- the
H$\alpha$ less than the bluer H$\beta$ and [OII] --, and by changes in
the upper end mass of the stellar Initial Mass function (IMF, which
affects the number of ionizing photons available to excite the
gas). The intensity of the Balmer lines is additionally affected by
the stellar absorption of the underlying galaxy stellar population --
again the H$\alpha$ at a lower level than H$\beta$. In contrast, the
[OII] is affected by the gas metallicity and, potentially, by its
ionization conditions (but, see, Kewley et~al.\ 2004, who find no such
influence for galaxies with O/H$>$8.5). A non--exhaustive list of
studies addressing such effects and/or deriving calibrations for SFR
estimates from line measurements includes: Gallagher et~al.\ 1989,
Kennicutt 1992, 1998, Charlot et~al.\  2002, Rosa--Gonzales et~al.\ 
2002, Kewley et~al. 2002, 2004, Perez--Gonzalez et~al.\ 2003,
Hopkins 2004.

Combined, the above effects impact SFR estimates from factors of a few
to orders of magnitude, depending on the regime where the measurement
is performed. This is easily seen in the case of dust
extinction. Local starburst galaxies cover a wide range of dust
attenuation values; even UV--selected starbursts can show extinctions
as high as A$_V\sim$4.5~mag, with a loose trend for more actively
star--forming galaxies to have higher extinctions (Figure~1, left;
see, also, Wang \& Heckman 1996, Sullivan et~al.\ 2001, Perez--Gonzalez
et~al.\ 2003). If such a highly extincted galaxy is mistakenly
corrected for a lower extinction value, e.g., A$_V\sim$1~mag (the value
derived for disks, Kennicutt 1983), the derived SFR(H$\alpha$) will
underestimate the actual one by a factor of 14! 

How common are such galaxies? From the H$\alpha$ luminosity function
of Perez--Gonzalez et~al.\ (2003), an L$^*$(H$\alpha$) galaxy in the
local Universe has A$^*_V\approx$2--3~mag, thus galaxies with heavily
extincted ionized gas are not a rarity in our cosmological
neighbourhood. How this translates to higher redshifts is still
unclear, although there are suggestions that overall extinctions are
decreasing at constant SFR in high-redshift galaxy populations
(Adelgerber \& Steidel 2000).

The SFR--extinction correlation carries down to the `quiescent'
star-forming galaxy regime, with a trend that is resemblant of the
starburst galaxies' one (Figure~1, right). The fact that more actively
star--forming galaxies and regions have, on average, larger dust
extinction values is a consequence of the Schmidt-Kennicutt law
(galaxies/regions with larger gas surface densities show larger
specific SFRs, e.g., Kennicutt 1998) combined with the
mass--metallicity relation (see, also, Tim Heckman's contribution to
this conference).

\begin{figure}[t]
\includegraphics[bb=0 350 300 650, width=1.2in, height=1.2in]{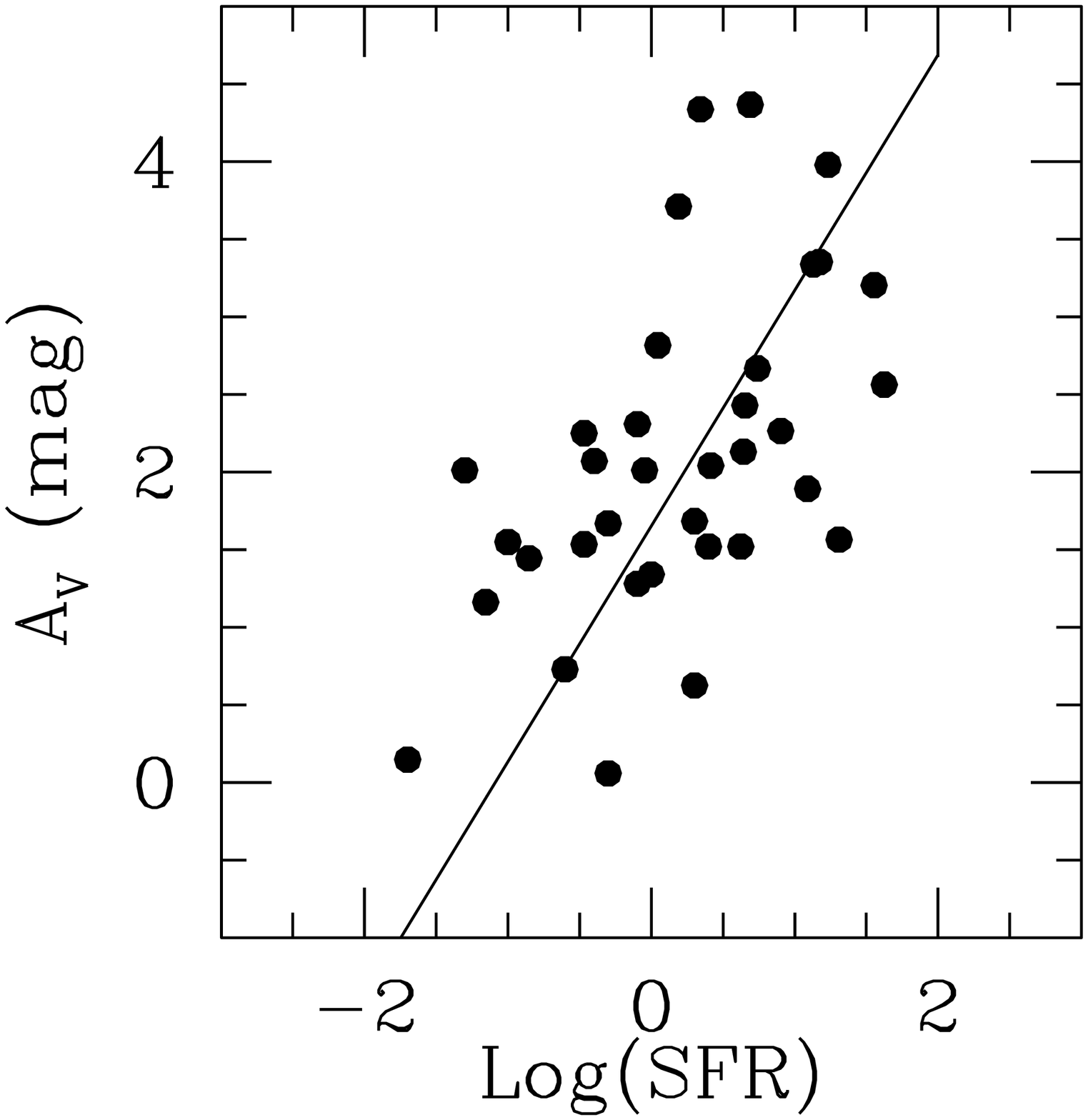}
\includegraphics[bb=-250 350 50 650, width=1.2in, height=1.2in]{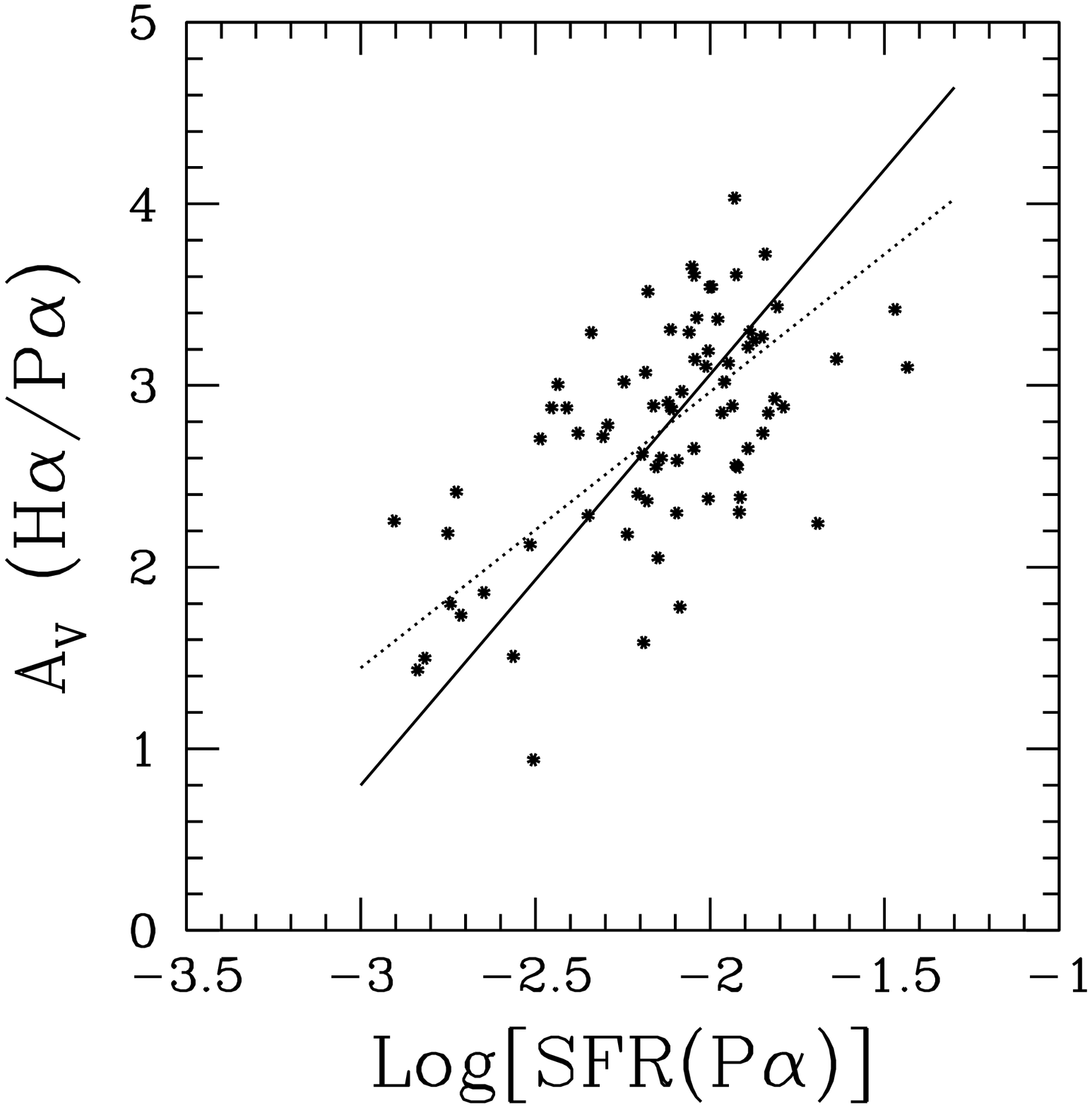}
\vskip0.8in
\caption{{\bf (Left)} The dust attenuation at V for the ionized gas,
A$_V$, versus SFR for the starburst galaxies sample of Calzetti et~al.\ 
(1994). The SFR is calculated from the extinction--corrected
H$\alpha$, while A$_V$ is derived from various hydrogen emission
lines, from H$\beta$ in the optical to Br$\gamma$ in the infrared. The
continuous line is the best linear fit through the datapoints. {\bf
(Right)} The A$_V$ versus SFR for HII-emitting complexes in the
central $\sim$6~kpc of M51a (Calzetti et~al.\ 2005). A$_V$ is derived
from the H$\alpha$/P$\alpha$ line ratio, and SFR is calculated from
the extinction--corrected P$\alpha$($\lambda$1.8756~$\mu$m). The
continuous line is the best fit through the points, while the dotted
line is the starburst galaxies fit (left panel), vertically rescaled
by a factor 10$^{4.35}$.}
\end{figure}

Measurements of the stellar IMF are ridden with controversy (see review
by Brandl \& Andersen 2004). Amid all this, the high-end of the
stellar IMF appears to be fairly constant from galaxy to galaxy in the
local Universe, and, possibly, has not changed from the high redshift
to the present day (Stasinska \& Leitherer 1996; Wyse et~al.\ 2002;
Elmegreen, this Conference). If variations do exist, they are likely a
second--order effect at least for what concerns SFR measurements. The
comparison between SFR(IR) and SFR(H$\alpha$) for a sample of local
galaxies by Kewley et~al.\ (2002) supports such statement. SFR(IR) is
the star formation calculated from the galaxies' far--infrared
emission, which is due to dust heated mainly by the stellar
non--ionizing radiation (UV, optical, etc.). Thus, two different
aspects of the stellar bolometric luminosity, the non--ionizing and
the ionizing ones, determine SFR(IR) and SFR(H$\alpha$),
respectively. The tight agreement between SFR(IR) and SFR(H$\alpha$),
within $\sim$10\%, over 4 orders of magnitude (Kewley et~al.\ 2002) is
supportive of a relatively constant upper-end of the IMF.

The impact of the underlying stellar absorption can be quite
significant on the Balmer emission lines, for two reasons: 1)~since
EW$_{abs}$(H$\alpha$)$\approx$EW$_{abs}$(H$\beta$), the intrinsically
weaker H$\beta$ line emission will be proportionally more `depressed'
by the underlying absorption than the H$\alpha$, altering measurements
of dust reddening; 2)~in galaxies where the star formation intensity
is proportionally a small fraction of the overall stellar emission,
the EW$_{abs}$(H$\alpha$) can be a significant fraction of the
EW$_{em}$(H$\alpha$), thus leading to underestimates of the emission
fluxes. Measurements of local star--forming galaxy populations
indicate values of EW$_{abs}$(H$\alpha$)$\sim$3--6~\AA\ (Kennicutt
1992, Calzetti et~al.\ 1994, Charlot et~al.\ 2002, Rosa-Gonzalez et~al.\ 
2002), with the smaller value more commonly applicable to starburst
galaxies.

These studies demonstrate that once the effects of dust extinction and
underlying stellar absorption are controlled, the H$\alpha$ emission 
is a reliable indicator of instantaneous SFR, and existing
calibrations (Kennicutt 1998) are sufficiently accurate for most
purposes. For SFR([OII]), there are additional effects to consider:
the dependence on metallicity and, potentially, on the ionization
conditions, but these can be `calibrated' in samples of local galaxies
(Kewley et~al.\ 2004) or empirical approaches can be adopted (Kennicutt
1998, Rosa-Gonzales et~al.\ 2002).  Problems arise when only a limited
amount of information, e.g., just one or two adjacent emission lines,
is available, as often is the case for samples at cosmological
distances. In these cases, the unknown dust extinction and underlying
stellar absorption corrections can lead to two effects: 1)~underestimates 
of the actual SFRs by a factor of a few, up to an order
of magnitude, depending on the nature of the sample and the restframe
wavelength region investigated; 2)~increase in the dispersion of the
SFR distribution of the sample by at least a factor of 2
(Rosa--Gonzales et~al.\ 2002).

\section{Star Formation Rate Indicators at Other Wavelengths}

The optical SFR indicators discussed above measure the presence and
amount of ionizing stars in galaxies, which are typically short--lived
(t$_{life}<$10$^7$~yr). Some of the other widely used SFR indicators
at shorter or longer wavelengths probe star formation over longer
timescales, typically 100~Myr or longer.

This is the case, for instance, of the restframe UV emission (where UV
is intended here as the stellar emission between 1000~\AA\ and
3000~\AA). While the UV emission is well correlated with the ionized
line emission in starburst galaxies (and thus both represent reliable
tracers of current SFR), there is increasing evidence that the UV of
quiescent star--forming galaxies traces recent, but not current, star
formation, and cannot be directly used as an `instantaneous' SFR
indicator (Kong et~al.\ 2004, Calzetti et~al.\ 2005).

Another popular SFR indicator is the one derived from a galaxy's far
infrared emission, SFR(IR). As mentioned in the previous section, the
far infrared emission in a galaxy is from dust heated mainly by the
non--ionizing stellar radiation. One of the standing questions is how
much of the IR radiation from a galaxy is contributed by current or
recent star formation, and how much by the general stellar radiation
field. The answer to this question can affect the empirical
calibration of SFR(IR).

Addressing some of those questions is one of the purposes of projects
like SINGS (the Spitzer Infrared Galaxies Survey, one of the Spitzer
Legacy Projects; P.I.: R.~Kennicutt; see J.D.~Smith's contribution to
this conference). The unprecedented angular resolution afforded by
Spitzer together with observations at multiple wavelengths of local
galaxies is enabling this project to shed light on the merits and
problems of common (and uncommon) SFR indicators, for use on galaxies
at cosmological distances.

\end{document}